\documentclass[usenatbib,usegraphicx]{mn2e}

\newcommand{\hI}{\textrm{H}\,\textsc{i}}
\newcommand{\mhI}{\mathrm{H\,I}}
\newcommand{\bootes}{Bo\"{o}tes}
\newcommand{\vlsr}{{\ensuremath{v_{\mathrm{LSR}}}}}
\newcommand{\kms}{{\ensuremath{\mathrm{km~s^{-1}}}}}
\newcommand{\Msun}{{\ensuremath{\mathrm{M}_{\sun}}}}
\newcommand{\feh}{{\ensuremath{\mathrm{[Fe/H]}}}}
\newcommand{\IHa}{{\ensuremath{I_{\mathrm{H}\alpha}}}}

\title[Absence of \hI\ in Boo dSph]{Absence of H\,I in the
 \bootes\ Dwarf Spheroidal Galaxy}

\author[Bailin \& Ford]{Jeremy Bailin$^1$ \& Alyson Ford$^{1,2}$\\
$^1$Centre for Astrophysics and Supercomputing,
 Swinburne University of Technology, Mail H39, PO Box 218, Hawthorn, Victoria,\\
 3122, Australia; jbailin@astro.swin.edu.au\\
$^2$Australia Telescope National Facility, CSIRO, P. O. Box 76, Epping NSW 1710, Australia}
\date{Accepted 2006 November 21.  Received 2006 November 14; in original form 2006
October 24}

\begin{document}

\maketitle

\begin{abstract}
Neutral hydrogen (\hI) observations towards the \bootes\ dwarf
spheroidal (dSph) galaxy (hereafter \bootes),
a very low luminosity metal-poor Galactic satellite,
were obtained using the Parkes Radio Telescope. We do not detect any \hI\
in or around \bootes\ to a $3\sigma$ upper limit of $180~\Msun$
within the optical half light radius and $8000~\Msun$ within 1.6~kpc.
Its \hI\ mass-to-light ratio is
less than $0.002~\Msun/L_{\sun}$, making \bootes\ one of the most
gas-poor galaxies known.
Either reionisation severely inhibited gas infall onto the proto-\bootes, or
large amounts of gas have been removed by ram pressure and/or tidal
stripping. Since \bootes\ lies on the mass-metallicity fundamental
line, this relation and the inefficiency of
star formation at the faintest end of the galaxy luminosity function
must be partly driven, or at least not disrupted, by extreme gas loss
in such low luminosity galaxies. We also do not detect any \hI\
associated with the leading tidal tail of the Sagittarius dSph
galaxy, which fortuitously passes through the
observed field, to a $3\sigma$ column density limit of $2\times
10^{17}~\mathrm{cm^{-2}}$.
This suggests that either the leading gaseous tail is ionised, or the
gas in the trailing tail was removed before the current
tidal disruption of the parent dSph began. 
  
\end{abstract}

\section{Introduction}\label{intro section}
Analysis of the stellar distribution in the Sloan Digital Sky Survey
has led to the recent discovery of many very faint dwarf satellite
galaxies of the Milky Way
\citep{willman-etal05-uma,zucker-etal06-cvn,zucker-etal06-uma2,
  belokurov-etal06-boo,belokurov-etal06-quintet,grillmair06-60deg}. 
Among the least luminous is the $M_V = -5.7$ satellite in \bootes\,
that was discovered by \citet{belokurov-etal06-boo} and has an
estimated distance of 60~kpc, with a right ascension (RA) of
$14^{\mathrm{h}}00^{\mathrm{m}}06^{\mathrm{s}}$ and declination of
$+14\degr 30\arcmin 00\arcsec$ ($l=358\fdg1$ and $b=69\fdg6$ in
Galactic coordinates). Its central surface brightness is
$\mu_{0,V}\approx 28~\mathrm{mag~arcsec^{-2}}$ with an optical
half-light radius of 13\arcmin. \citet{munoz-etal06} obtained spectra
of a number of red giant branch and asymptotic giant branch stars in
\bootes\ and found a heliocentric systemic velocity of $95.6~\kms$ (or
$\vlsr=106.2~\kms$, where \vlsr\ is the velocity with respect to the
Local Standard of Rest) and a velocity dispersion of $6.6~\kms$. This
implies a dynamical mass of $\approx 10^7~\Msun$, similar to that of most
low luminosity dwarfs \citep{mateo98}. The combined spectra of \bootes\ 
member stars are consistent with a metallicity of $\feh \sim -2.5$,
making it the most metal-poor galaxy known in the Local Group. From the
periods of 15~RR~Lyrae variable stars in \bootes, \citet{siegel06} also
calculated a metallicity of $\feh \sim -2.5$ and a distance of
$62\pm4~\mathrm{kpc}$.  

Based on the population of red giant branch stars in its colour
magnitude diagram, \citet{munoz-etal06} suggested that the luminosity
of \bootes\ may be a factor of almost five
larger than the estimate of
\citet{belokurov-etal06-boo}, implying $M_V = -7.5$. \citet{siegel06} also
argued for a high luminosity based on the population of RR~Lyrae stars.
Further support for this higher luminosity comes from
the ``fundamental line'', a tight correlation between the mass-to-light
ratio, metallicity, and central surface brightness of Local Group
satellite galaxies \citep{pb02}.
Assuming $\feh = -2.5$ and
$\mu_{0,V} = 28~\mathrm{mag~arcsec^{-2}}$, this relation predicts a
dynamical mass-to-light ratio for \bootes\ of $M/L_V \approx
150~\Msun/L_{\sun}$, in
agreement with the higher luminosity estimate of
\citet{munoz-etal06}, which we therefore adopt. The continuation of
the fundamental line down to the most metal-poor galaxy in the Local
Group makes \bootes\ an ideal place to probe the origin of the relation.

Although the environs of the Milky Way contain a number of low-mass
faint dwarf galaxies, cosmological simulations predict an even larger
number of potential host subhaloes
\citep{klypin-etal99,moore-etal99}. A natural interpretation for this
discrepancy is that lower-mass haloes are increasingly inefficient at
converting their primordial baryons into stars. This interpretation is
bolstered by the internal kinematics of the lowest luminosity dwarfs, whose
luminosities drop dramatically as they approach a universal halo mass
of $10^7~\Msun$, suggesting that this is the critical mass for efficient
star formation \citep{mateo98}. 

Gas loss from supernova feedback and stellar winds have been invoked to
explain both the low luminosities of low-mass galaxies and the
mass-metallicity relation \citep{dw03,rpv06}. In the vicinity of
larger galaxies, gas can also be lost to ram pressure stripping
\citep{br00,ggh03}, tidal stripping \citep{ckg06}, or both
\citep{mayer-etal06}. In contrast, hydrodynamic simulations of dwarf
galaxies by \citet{tkg06} and \citet{brooks-etal06-mzr} suggest that
the low star formation efficiency is driven not by gas loss but by
self-regulated star formation consistent with the \citet{kennicutt98}
star formation law. Reionisation may also dramatically
reduce the number of baryons available to form stars in many low-mass
haloes \citep{kbs97,bkw00,dlm03}. 

The predicted properties of the interstellar media of dwarf galaxies
are different in each of these scenarios. If dwarf galaxies are
inefficient at converting their available baryons into stars
then they will still contain large reservoirs of gas.
On the other hand, tidal stripping leaves streams of gas
extending both in front of and behind dwarfs, while ram pressure
stripping leaves only trailing streams.
If reionisation inhibits gas infall then there will be no discernible
gas associated with dwarfs.

\citet{br00} examined the Leiden-Dwingeloo \hI\ Survey
(\citealt{leiden-dwingeloo-survey}, LDS; see also \citealt{lab-survey})
for \hI\
emission associated with most Local Group dSph
galaxies. They found
that many dSphs contain large amounts of gas, with typical
velocity widths of $20~\kms$, but that those within 250~kpc of either
the Milky Way or M31 are devoid of \hI\ (see also
\citealp{bl99}).
They argued that ram pressure
stripping is responsible for this deficit of gas among nearby satellites.

The relative proximity of the \bootes\ dSph galaxy
allows us to detect very small quantities of gas and probe the
relationship between stellar mass, gas mass and dynamical mass at the
faintest end of the galaxy luminosity function. In this paper we
present \hI\ observations in the direction of \bootes, obtain strict
upper limits for its \hI\ mass, and use those limits to constrain the
origin of its extremely low luminosity. By coincidence, the leading
stellar tidal tail of the Sagittarius dSph galaxy, another
satellite of the Milky Way which is in the process of tidal
disruption, passes in front of \bootes\ along the line of sight
\citep{munoz-etal06}. \hI\ has been identified in the trailing tail
\citep{putman-etal04}, but none has yet been detected in the leading
tail as might be expected if the gas were tidally stripped along with
the stars. We obtain strict upper limits on the column density of \hI\
associated with the Sagittarius leading tail.

\section{Observations and Data Reduction}\label{observations section}

Observations were obtained at the 64m Parkes Radio Telescope in
June~2006 using the 20cm multibeam receiver
\citep{parkes-multibeam}. The 13 beams of the receiver are arranged in
a hexagonal pattern with a 96\arcmin-wide footprint. To ensure full
spatial sampling within the inner  $\sim$2\fdg5 $\times$ 2\fdg5 of the
field and to reduce artefacts introduced by scanning, scans in both
constant RA and declination were observed, at a rate of
$1\degr~\mathrm{min^{-1}}$. Each scan was 2\degr9\arcmin\ in length
and adjacent scans were offset by 32\arcmin. An approximately 
3\degr $\times$ 3\degr\ field was mapped, with 33 sets of 3 RA scans
interleaved with 32 sets of 3 declination scans, or 195 individual
scans total. To provide wider spatial coverage, these data were
combined with 9 declination scans 8\degr\ in length spanning an area
approximately 6\degr $\times$ 8\degr\ centred on \bootes.

The receiver was operated in frequency switching mode with a throw of
3.125~MHz centred on 1420.4015~MHz, with 2048 channels over an 8~MHz
bandwidth, resulting in a velocity coverage of $-400 \le \vlsr \le
450$~\kms\ and a channel width of $0.8~\kms$.
The spatial resolution of the data is $15\arcmin$.
Fluxes were calibrated
from observations of the standards S6, S8, and S9 \citep{williams73}.
Although the data have not been corrected
for stray radiation, it is unlikely to impact our observations due to
the high Galactic latitude of the observed field.    

\begin{figure*}
\includegraphics[scale=0.45]{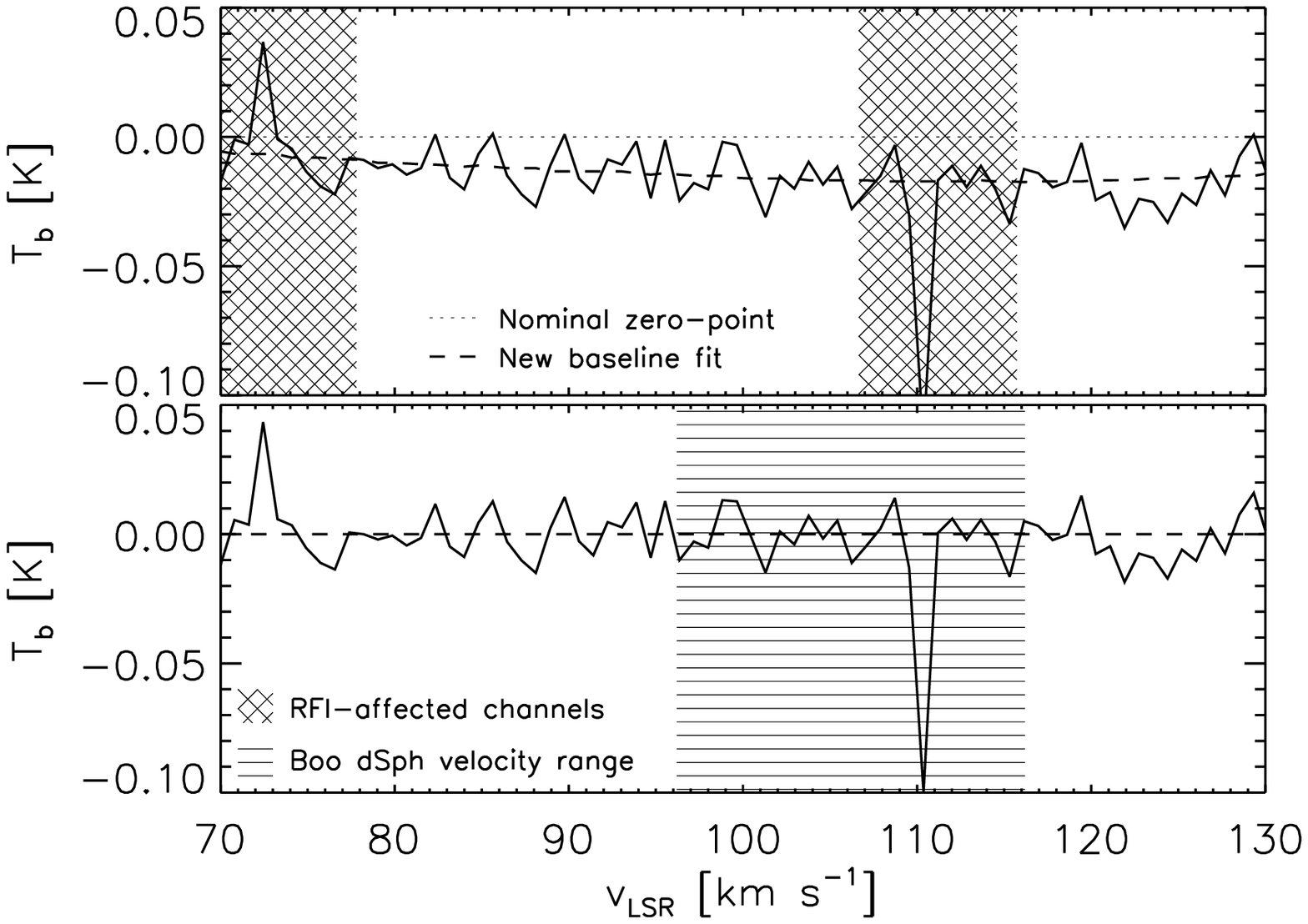}\hspace{1.2cm}
\includegraphics[scale=0.45]{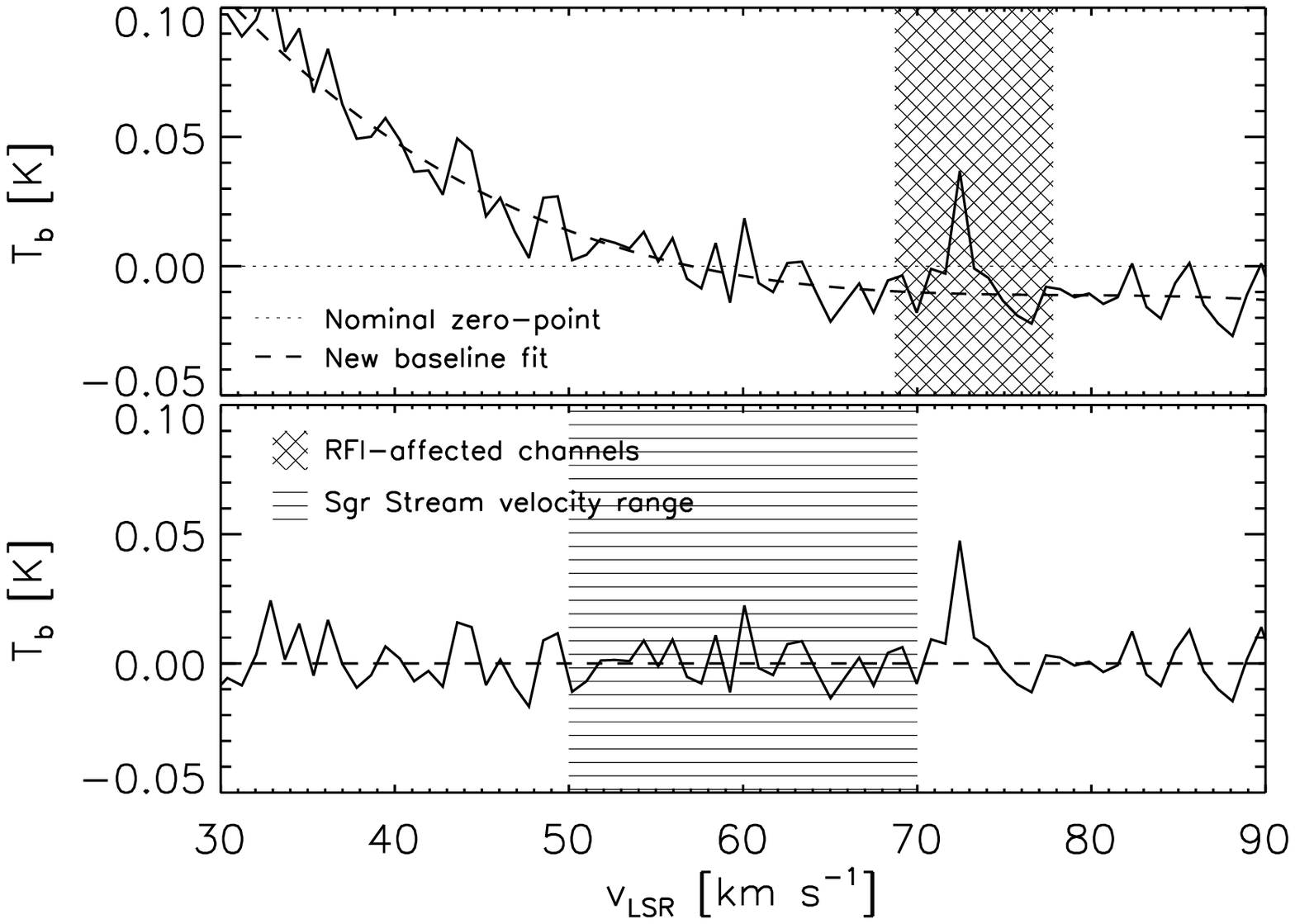}
\caption{\label{continuum fit figure}%
\textit{(Top-left)} Spectrum at the location and velocity of \bootes\ 
after initial data reduction. The dotted line indicates
the nominal zero-point while the dashed line shows
our adopted baseline.
The cross-hatched region indicates channels suffering from low-level RFI.
\textit{(Bottom-left)} As above but corrected for the new baseline.
The shaded region shows a $20~\kms$ velocity range around the
nominal \bootes\ velocity.
\textit{(Right)} As in the left panels but at the velocity expected
for the Sagittarius leading tidal tail.
Local Galactic emission is visible at the lower velocities.}
\end{figure*}

Raw data were reduced using Livedata and then gridded into a 3D cube
using Gridzilla, where
data from beam 12 were omitted due to a failure of one of its low noise
amplifiers. All analysis of the data cube was performed using IDL.
After bandpass calibration, baseline residuals at the 10~mK level
were present;
we therefore corrected for these residuals
by determing the baseline on a pixel-by-pixel basis using a median
filter of width $50~\kms$. This provided a good baseline fit except
at $-60 \le \vlsr \le 60~\kms$, where Galactic emission biases the
median.
Therefore, when analysing the data at the expected Sagittarius
stream velocity \citep{munoz-etal06}, we fitted a 3rd-degree
polynomial over a 50~\kms\ wide velocity range surrounding
$\vlsr=60~\kms$.
Examples of this process are shown in Figure~\ref{continuum fit figure}.

\begin{figure}
\includegraphics[scale=0.45]{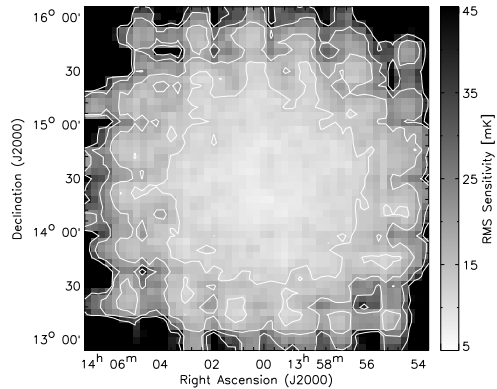}
\caption{\label{sensitivity map}%
Map of the rms sensitivity achieved in each 0.8~\kms\ channel.
Contours are drawn at 12.5, 17.5, 25, and 35~mK.}
\end{figure}

The rms brightness temperature sensitivity achieved in each channel
varies across the field due to the arrangement of the beams, with a
maximum sensitivity of 7~mK and a median
sensitivity of 13~mK within 1\fdg5 of the field centre. The
sensitivity was empirically determined  for each pixel from the rms
dispersion in an emission-free region of the spectrum after baseline
subtraction, and is shown in Figure~\ref{sensitivity map}.  

Examination of individual channel maps reveals that the data over
velocity ranges of $68.7 \le \vlsr \le 77.8$~\kms\ and $106.7 \le
\vlsr \le 115.7$~\kms\ are affected by low-level radio frequency
interference (RFI) of unknown origin. We mask out all affected
channels in the analysis and discuss the effects on our upper limits
in \S~\ref{boo results}. 

\section{Results for Boo dSph}\label{boo results}

\begin{figure}
\includegraphics[scale=0.45]{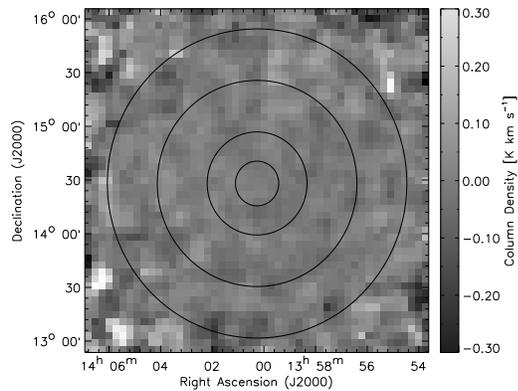}
\caption{\label{mom0 boo figure}%
Zeroth moment map at the velocity of \bootes.
Apertures of radius 13\arcmin\ (the optical half-light radius),
30\arcmin, 60\arcmin, and 90\arcmin\ are overplotted.
The noise increases at the edges of the image
due to the decreased integration time (see Figure~\ref{sensitivity map}).}
\end{figure}

We find no evidence of \hI\ emission at the velocity of \bootes. In
Figure~\ref{mom0 boo figure} we present the zeroth moment map
integrated over a velocity width of $20~\kms$ centred at
$\vlsr=106.2~\kms$, excluding RFI-affected channels. No
statistically significant features are
seen over the entire 3\degr $\times$ 3\degr\ field (3.5 x 3.5~kpc at a
distance of 60~kpc). We have calculated the integrated line flux
both at the central pixel and in a
number of circular apertures with radii ranging
from the half light radius of 13\arcmin\ 
to 90\arcmin. Sample averaged spectra are shown in Figure~\ref{boo
  aperture spec figure}, while corresponding mean column densities,
$\left<N_{\mhI}\right>$, and \hI\ masses, $M_{\mhI}$, are given in
Table~\ref{boo line flux table}. The quoted $1\sigma$ uncertainty has been
calculated empirically by performing an identical analysis at 200
randomly-chosen central velocities located far from local Galactic
emission and taking the dispersion of these measurements; this method
implicitly takes into account systematics such as residual
baseline-fitting errors in addition to the random error component.

\begin{figure}
\includegraphics[scale=0.45]{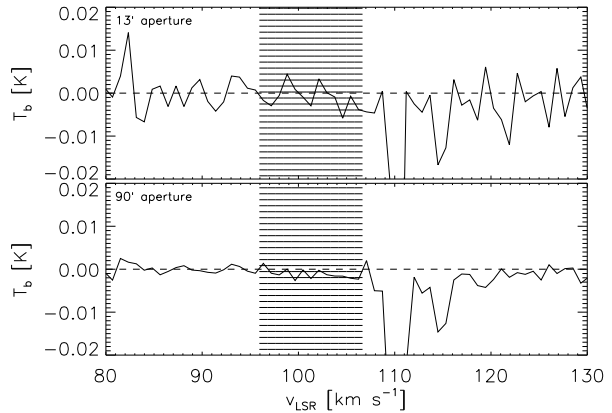}
\caption{\label{boo aperture spec figure}%
Spectra averaged over a 13\arcmin\ \textit{(top)} and 
90\arcmin\ \textit{(bottom)} radius aperture centred on \bootes.
The velocity range of the zeroth moment map (Figure~\ref{mom0 boo figure})
is represented by the shaded region.}
\end{figure}

\begin{table}
\caption{Boo dSph Line Fluxes}
\label{boo line flux table}
\begin{tabular}{@{}lccc}
\hline
Aperture & Integrated Line Flux & $\left<N_{\mhI}\right>$ & $M_{\mhI}$ \\
Radius & [K~\kms] & [$10^{16}~\mathrm{cm^{-2}}$] & [\Msun]\\
\hline
1~pixel & $-0.010\pm0.028$ & $-1.7\pm5.1$ & $-7\pm20^a$\\
13\arcmin & $-0.010\pm0.014$ & $-1.7\pm2.5$ & $-23\pm33$\\
30\arcmin & $-0.017\pm0.012$ & $-3.0\pm2.1$ & $-220\pm150$\\
60\arcmin & $-0.013\pm0.012$ & $-2.4\pm2.2$ & $-670\pm620$\\
90\arcmin & $-0.012\pm0.012$ & $-2.3\pm2.2$ & $-1400\pm1400$\\
\hline
\end{tabular}
$^a$Equivalent mass within the beam half power radius.
\end{table}

Taking a $3\sigma$ upper limit, we conclude that there is no more than
$100~\Msun$ of \hI\ at velocities $96 \le \vlsr \le 107~\kms$ within
the optical half-light radius of \bootes. Although we cannot measure
an upper limit in the region of the spectrum affected by RFI, we can
rule out any significant reservoir of \hI, which would be evident in the clean
part of the spectrum. If we assume that any emission at RFI-affected
velocities is of similar magnitude to that in the clean part of the
spectrum then the upper limit on the total \hI\ mass within the
optical half-light radius is $180~\Msun$. This assumption is likely to
be valid unless the \hI\ emission is much narrower than the emission
seen in other dSphs \citep{br00} and offset to
positive velocity with respect to the stars, a possibility we consider
unlikely as gas removed by ram pressure stripping would trail the
stars and appear offset to negative velocity. In an aperture with
radius 90\arcmin\ (1.6~kpc at a distance of 60~kpc)
we find a $3\sigma$ upper limit of
$4300~\Msun$ of \hI\ at $96 \le \vlsr \le 107~\kms$, and a total
RFI-corrected \hI\ mass upper limit of $8000~\Msun$. The $3\sigma$
column density limit within the beam is $1.5\times
10^{17}~\mathrm{cm^{-2}}$ at $96 \le \vlsr \le 107~\kms$ and
$2.8\times 10^{17}~\mathrm{cm^{-2}}$ correcting for RFI-affected
channels.

\section{Results for Sgr Leading Tail}\label{sgr results}

\citet{putman-etal04} found no detections of positive velocity 
\hI\ along the orbit of the Sagittarius dSph in
the \hI\ Parkes All-Sky Survey (HIPASS).
The HIPASS observations north of the celestial equator
reach an rms sensitivity of 11~mK with a spectral resolution
of $26.4~\kms$, corresponding to a
column density $3\sigma$ upper limit of $1.6\times 10^{18}~\mathrm{cm^{-2}}$.

\begin{figure}
\includegraphics[scale=0.45]{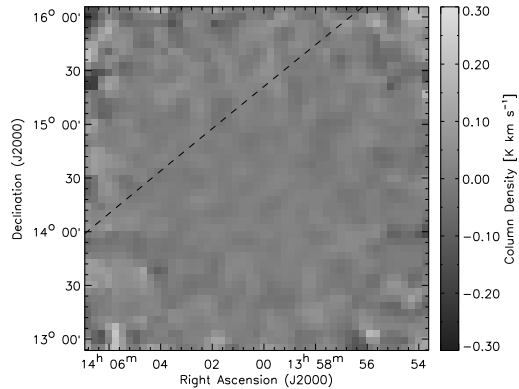}
\caption{\label{mom0 sgr figure}%
Zeroth moment map at the velocity of the Sagittarius leading tidal tail.
The dashed line indicates $l=0\degr$. Features associated with Sagittarius
are expected to lie parallel to this line.
The noise increases at the edges of the image
due to the decreased integration time (see Figure~\ref{sensitivity map}).}
\end{figure}

\begin{figure}
\includegraphics[scale=0.45]{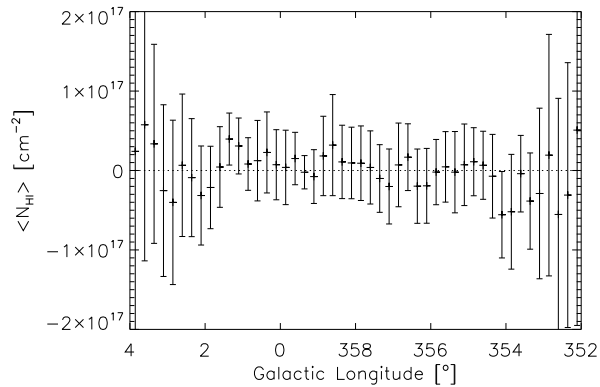}
\caption{\label{sgr l strip figure}%
Mean \hI\ column density in strips of constant Galactic longitude
at the velocity range of Sagittarius debris. Strips are $\Delta l=15\arcmin$
wide and include the region shown in Figure~\ref{mom0 sgr figure}.}
\end{figure}

In Figure~\ref{mom0 sgr figure} we present our zeroth moment map
integrated over a velocity width of $20~\kms$ centred at
$\vlsr=60~\kms$.
These much deeper observations still reveal no
evidence for any emission at the velocity of the Sagittarius leading tidal
tail. Due to the polar orbit of Sagittarius, stream gas is expected to lie
nearly parallel to lines of constant Galactic longitude, where
$l=0\degr$ is indicated by the dashed line in Figure~\ref{mom0 sgr
  figure}. No features are found elongated in this direction. The
column density $3\sigma$ upper limit averaged over the central
$2\times 2$ pixels is $2\times 10^{17}~\mathrm{cm^{-2}}$. In an
attempt to enhance any faint signal, we have calculated the mean
column density in strips of constant $l$ with widths, $\Delta l$,
ranging from 15\arcmin\ to 90\arcmin. An example for strips of width
$\Delta l = 15\arcmin$, corresponding closely to the beam FWHM,
is shown in Figure~\ref{sgr l strip figure} and does not
reveal any emission. We conclude that there is no \hI\ associated with
the leading Sagittarius tidal tail at this location down to a $3\sigma$ column
density limit of $2\times 10^{17}~\mathrm{cm^{-2}}$.

The leading tail also appears to be devoid of ionised gas, based on the
non-detection of H$\alpha$ emission at this location and velocity in the
Wisconsin H-Alpha Mapper Northern Sky Survey
(WHAM-NSS; \citealt{wham-nss}) to a $3\sigma$ intensity upper limit
of $\IHa < 0.025~\mathrm{R}$. If ionised leading stream gas exists
with the same column density as the neutral gas
that is in the trailing tail, then its volume density must be less than
$\bar{n}_{\mathrm{H}} < 0.002~\mathrm{cm^{-3}}$,
implying an unreasonably large line-of-sight
depth of $L > 17~\mathrm{kpc}$. However, a moderate reduction in the
mean column density within the $1\degr$ WHAM beam could lower $L$
to plausible values.

\section{Conclusions}

We detect no \hI\ gas associated with the \bootes\ dSph
galaxy down to a
$3\sigma$ column density limit of $2.8\times 10^{17}~\mathrm{cm^{-2}}$,
corresponding to a mass limit of $180~\Msun$ within
the optical half-light radius and $8000~\Msun$ within 1.6~kpc of the
optical galaxy. While the stellar body of \bootes\ appears distorted
\citep{belokurov-etal06-boo},
casting doubt on its morphological classification,
the dearth of gas in \bootes\ clearly indicates that it is an early-type
dSph galaxy.

\begin{figure}
\includegraphics[scale=0.45]{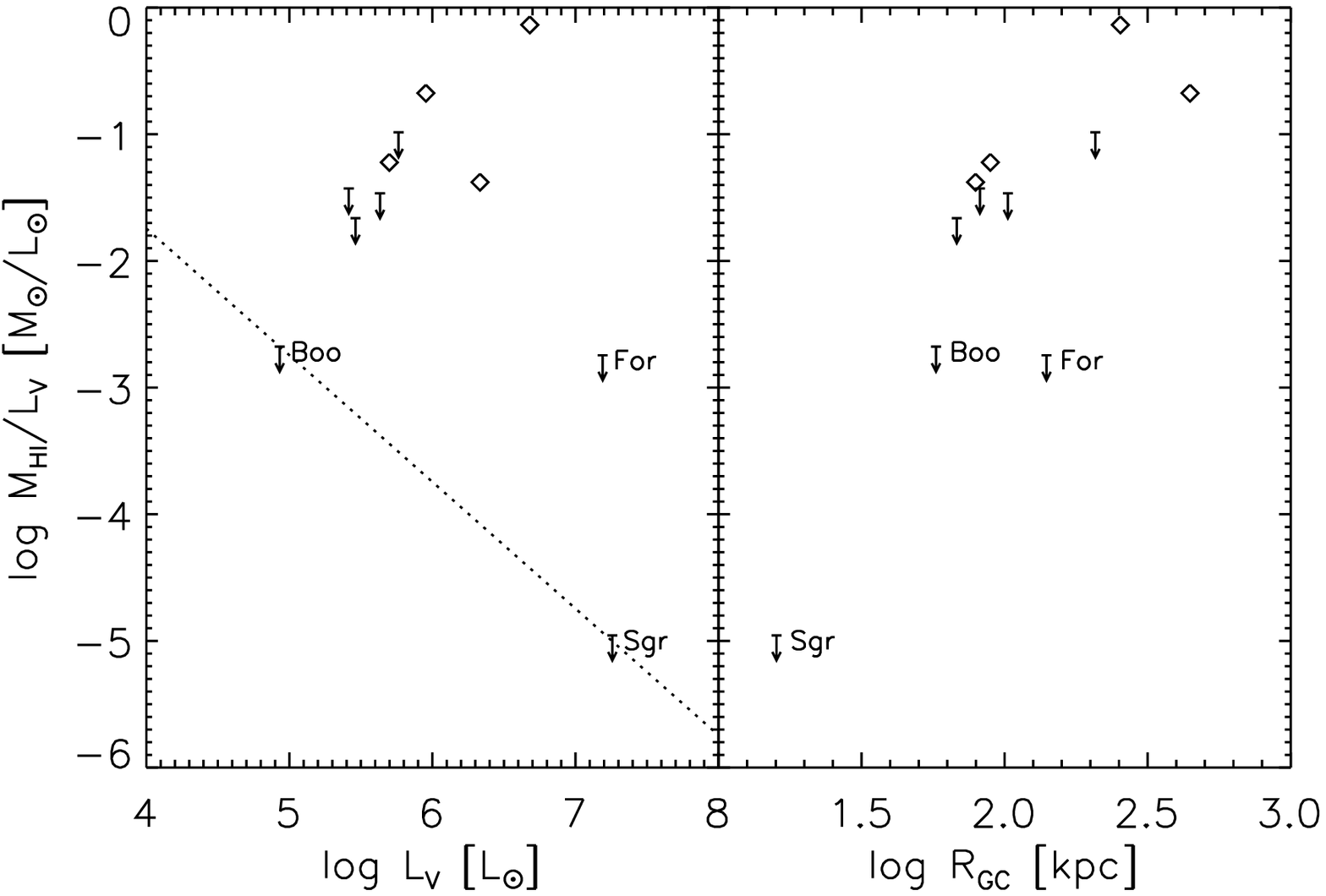}
\caption{\label{HI frac figure}%
Gas fraction
(ratio between \hI\ gas mass and V-band luminosity) of
dSph satellite galaxies
of the Milky Way as a function of their V-band luminosity,
$L_{\mathrm{V}}$, \textit{(left)},
and galactocentric radius, $R_{\mathrm{GC}}$, \textit{(right)}.
The upper limit on the total \hI\ mass of \bootes\ within the optical
half-light radius is $180~\Msun$ (\S~\ref{boo results}) and
the upper limit for Sagittarius is $200~\Msun$ \citep{bl99}.
All other \hI\ masses and upper limits are from
\citet{br00}.
A dotted line of constant \hI\ mass is shown in the left panel.
Exceptionally gas-poor galaxies are labelled.}
\end{figure}

In Figure~\ref{HI frac figure} we plot the gas fraction of \bootes,
as estimated by the ratio of the \hI\ gas mass to the
V-band luminosity, along with the other dSph companions
to the Milky Way catalogued by \citet{br00}
(for Sagittarius we use the deeper limit of \citealt{bl99}).
Our observations are more than 5 times deeper than those of the LDS;
combined with the relative proximity of \bootes, we are able to detect
much smaller quantities of \hI\ than \citet{br00}.
\bootes\ has the most stringent upper limit on its absolute
\hI\ mass of any Local Group galaxy, and is among the most
gas-poor galaxies even relative to its extremely low luminosity,
with a $3\sigma$ upper limit of $M_{\mhI}/L_V < 0.002~\Msun/L_{\sun}$.
Only the much more luminous Sagittarius and Fornax dSphs
are known to be more gas-poor.

\bootes\ has converted only a small fraction of its primordial baryons
into stars.
The leftover gas is no longer part of \bootes\ as would be
expected if self-regulated star formation had been the only process
operating.
Unless reionisation prevented cool gas from falling onto \bootes\ 
at early times, gas must have been removed by
tidal forces, ram pressure stripping, or supernova feedback.
The proximity-gas mass relation (\citealp{br00}; Figure~\ref{HI frac figure})
argues strongly against supernova feedback as the primary mechanism,
since it occurs in all environments.
Neither streams of stripped gas nor outflowing gas
from \bootes\ are detected, indicating
that these gas removal processes have either already completed or
have left streams too tenuous to be detected with current observations.
As \bootes\ lies on the fundamental line \citep{pb02}, the explanation
for such a mass-metallicity relation cannot be entirely due to
inefficient self-regulated star formation but must be partially
driven, or at least not disrupted,
by the extreme gas loss or primordial deficit in cool gas
experienced by this low luminosity galaxy.

We detect no neutral gas associated with the leading Sagittarius tidal tail
down to a $3\sigma$ column density limit of 
$2\times 10^{17}~\mathrm{cm^{-2}}$.
If the gas detected along the trailing tail \citep{putman-etal04}
had been tidally stripped along with the stars, then a
leading tail of gas would also exist.
Either this leading tail is ionised, which would require it to
have a lower density than
the trailing tail, or the gas in the trailing tail
was removed by ram pressure stripping before the tidal
disruption of the parent galaxy began.

No other non-Galactic emission is detected at
$-400 \le \vlsr \le 450$~\kms\ over the entire
6\degr $\times$ 8\degr\ field.

\section*{Acknowledgments}
The Parkes Observatory is operated by the Australia Telescope National
Facility, a division of the Commonwealth Scientific and Industrial
Research Organisation.
The Wisconsin H-Alpha Mapper is funded by the National Science
Foundation.
We thank John Reynolds, Andrew Hunt, and Jay
Lockman for useful discussions.
JB thanks the Australian Research Council for financial support.   

\bibliography{../../masterref.bib}

\begin{thebibliography}{32}
\expandafter\ifx\csname natexlab\endcsname\relax\def\natexlab#1{#1}\fi

\bibitem[{{Belokurov} {et~al.}(2006{\natexlab{a}}){Belokurov}, {Zucker},
  {Evans}, {Wilkinson}, {Irwin}, {Hodgkin}, {Bramich}, {Irwin}, {Gilmore},
  {Willman}, {Vidrih}, {Newberg}, {Wyse}, {Fellhauer}, {Hewett}, {Cole},
  {Bell}, {Beers}, {Rockosi}, {Yanny}, {Grebel}, {Schneider}, {Lupton},
  {Barentine}, {Brewington}, {Brinkmann}, {Harvanek}, {Kleinman}, {Krzesinski},
  {Long}, {Nitta}, {Smith}, \& {Snedden}}]{belokurov-etal06-boo}
{Belokurov}, V. et~al. 2006{\natexlab{a}}, ApJ, 647, L111

\bibitem[{{Belokurov} {et~al.}(2006{\natexlab{b}}){Belokurov}, {Zucker},
  {Evans}, {Kleyna}, {Koposov}, {Hodgkin}, {Irwin}, {Gilmore}, {Wilkinson},
  {Fellhauer}, {Bramich}, {Hewett}, {Vidrih}, {De Jong}, {Smith}, {Rix},
  {Bell}, {Wyse}, {Newberg}, {Mayeur}, {Yanny}, {Rockosi}, {Gnedin},
  {Schneider}, {Beers}, {Barentine}, {Brewington}, {Brinkmann}, {Harvanek},
  {Kleinman}, {Krzesinski}, {Long}, {Nitta}, \&
  {Snedden}}]{belokurov-etal06-quintet}
{Belokurov}, V. et~al. 2006{\natexlab{b}}, ApJ, in press, astro-ph/0608448

\bibitem[{{Blitz} \& {Robishaw}(2000)}]{br00}
{Blitz}, L., \& {Robishaw}, T. 2000, ApJ, 541, 675

\bibitem[{{Brooks} {et~al.}(2006){Brooks}, {Governato}, {Booth}, {Willman},
  {Gardner}, {Wadsley}, {Stinson}, \& {Quinn}}]{brooks-etal06-mzr}
{Brooks}, A.~M., {Governato}, F., {Booth}, C.~M., {Willman}, B., {Gardner},
  J.~P., {Wadsley}, J., {Stinson}, G., {Quinn}, T. 2006, ApJL, submitted,
  astro-ph/0609620

\bibitem[{{Bullock} {et~al.}(2000){Bullock}, {Kravtsov}, \& {Weinberg}}]{bkw00}
{Bullock}, J.~S., {Kravtsov}, A.~V., {Weinberg}, D.~H. 2000, ApJ, 539, 517

\bibitem[{{Burton} \& {Lockman}(1999)}]{bl99}
{Burton}, W.~B., {Lockman}, F.~J. 1999, A\&A, 349, 7

\bibitem[{{Connors} {et~al.}(2006){Connors}, {Kawata}, \& {Gibson}}]{ckg06}
{Connors}, T.~W., {Kawata}, D., {Gibson}, B.~K. 2006, MNRAS, 371, 108

\bibitem[{{Dekel} \& {Woo}(2003)}]{dw03}
{Dekel}, A., {Woo}, J. 2003, MNRAS, 344, 1131

\bibitem[{{Dong} {et~al.}(2003){Dong}, {Lin}, \& {Murray}}]{dlm03}
{Dong}, S., {Lin}, D.~N.~C., {Murray}, S.~D. 2003, ApJ, 596, 930

\bibitem[{{Grebel} {et~al.}(2003){Grebel}, {Gallagher}, \& {Harbeck}}]{ggh03}
{Grebel}, E.~K., {Gallagher}, III, J.~S., {Harbeck}, D. 2003, AJ, 125, 1926

\bibitem[{{Grillmair}(2006)}]{grillmair06-60deg}
{Grillmair}, C.~J. 2006, ApJ, 645, L37

\bibitem[{{Haffner} {et~al.}(2003){Haffner}, {Reynolds}, {Tufte}, {Madsen},
  {Jaehnig}, \& {Percival}}]{wham-nss}
{Haffner}, L.~M., {Reynolds}, R.~J., {Tufte}, S.~L., {Madsen}, G.~J.,
  {Jaehnig}, K.~P., {Percival}, J.~W. 2003, ApJS, 149, 405

\bibitem[{{Hartmann} \& {Burton}(1997)}]{leiden-dwingeloo-survey}
{Hartmann}, D., {Burton}, W.~B. 1997, {Atlas of Galactic Neutral Hydrogen}
  (Cambridge University Press)

\bibitem[{{Kalberla} {et~al.}(2005){Kalberla}, {Burton}, {Hartmann}, {Arnal},
  {Bajaja}, {Morras}, \& {P{\"o}ppel}}]{lab-survey}
{Kalberla}, P.~M.~W., {Burton}, W.~B., {Hartmann}, D., {Arnal}, E.~M.,
  {Bajaja}, E., {Morras}, R., {P{\"o}ppel}, W.~G.~L. 2005, A\&A, 440, 775

\bibitem[{{Kennicutt}(1998)}]{kennicutt98}
{Kennicutt}, Jr., R.~C. 1998, ApJ, 498, 541

\bibitem[{{Kepner} {et~al.}(1997){Kepner}, {Babul}, \& {Spergel}}]{kbs97}
{Kepner}, J.~V., {Babul}, A., {Spergel}, D.~N. 1997, ApJ, 487, 61

\bibitem[{{Klypin} {et~al.}(1999){Klypin}, {Kravtsov}, {Valenzuela}, \&
  {Prada}}]{klypin-etal99}
{Klypin}, A., {Kravtsov}, A.~V., {Valenzuela}, O., {Prada}, F. 1999, ApJ,
  522, 82

\bibitem[{{Mateo}(1998)}]{mateo98}
{Mateo}, M.~L. 1998, ARA\&A, 36, 435

\bibitem[{{Mayer} {et~al.}(2006){Mayer}, {Mastropietro}, {Wadsley}, {Stadel},
  \& {Moore}}]{mayer-etal06}
{Mayer}, L., {Mastropietro}, C., {Wadsley}, J., {Stadel}, J., {Moore}, B.
  2006, MNRAS, 369, 1021

\bibitem[{{Moore} {et~al.}(1999){Moore}, {Ghigna}, {Governato}, {Lake},
  {Quinn}, {Stadel}, \& {Tozzi}}]{moore-etal99}
{Moore}, B., {Ghigna}, S., {Governato}, F., {Lake}, G., {Quinn}, T., {Stadel},
  J., {Tozzi}, P. 1999, ApJ, 524, L19

\bibitem[{{Mu\~{n}oz} {et~al.}(2006){Mu\~{n}oz}, {Carlin}, {Frinchaboy}, {Nidever},
  {Majewski}, \& {Patterson}}]{munoz-etal06}
{Mu\~{n}oz}, R.~R., {Carlin}, J.~L., {Frinchaboy}, P.~M., {Nidever}, D.~L.,
  {Majewski}, S.~R., {Patterson}, R.~J. 2006, ApJ, 650, L51

\bibitem[{{Prada} \& {Burkert}(2002)}]{pb02}
{Prada}, F., {Burkert}, A. 2002, ApJ, 564, L73

\bibitem[{{Putman} {et~al.}(2004){Putman}, {Thom}, {Gibson}, \&
  {Staveley-Smith}}]{putman-etal04}
{Putman}, M.~E., {Thom}, C., {Gibson}, B.~K., {Staveley-Smith}, L. 2004,
  ApJ, 603, L77

\bibitem[{{Read} {et~al.}(2006){Read}, {Pontzen}, \& {Viel}}]{rpv06}
{Read}, J.~I., {Pontzen}, A.~P., {Viel}, M. 2006, MNRAS, 371, 885

\bibitem[{{Siegel}(2006)}]{siegel06}
{Siegel}, M.~H. 2006, ApJ, 649, L83

\bibitem[{{Staveley-Smith} {et~al.}(1996){Staveley-Smith}, {Wilson}, {Bird},
  {Disney}, {Ekers}, {Freeman}, {Haynes}, {Sinclair}, {Vaile}, {Webster}, \&
  {Wright}}]{parkes-multibeam}
{Staveley-Smith}, L. et~al. 1996, PASA, 13, 243

\bibitem[{{Tassis} {et~al.}(2006){Tassis}, {Kravtsov}, \& {Gnedin}}]{tkg06}
{Tassis}, K., {Kravtsov}, A.~V., {Gnedin}, N.~Y. 2006, ApJ, submitted,
  astro-ph/0609763

\bibitem[{{Williams}(1973)}]{williams73}
{Williams}, D.~R.~W. 1973, A\&AS, 8, 505

\bibitem[{{Willman} {et~al.}(2005){Willman}, {Dalcanton}, {Martinez-Delgado},
  {West}, {Blanton}, {Hogg}, {Barentine}, {Brewington}, {Harvanek}, {Kleinman},
  {Krzesinski}, {Long}, {Neilsen}, {Nitta}, \& {Snedden}}]{willman-etal05-uma}
{Willman}, B. et~al. 2005, ApJ, 626, L85

\bibitem[{{Zucker} {et~al.}(2006{\natexlab{a}}){Zucker}, {Belokurov}, {Evans},
  {Wilkinson}, {Irwin}, {Sivarani}, {Hodgkin}, {Bramich}, {Irwin}, {Gilmore},
  {Willman}, {Vidrih}, {Fellhauer}, {Hewett}, {Beers}, {Bell}, {Grebel},
  {Schneider}, {Newberg}, {Wyse}, {Rockosi}, {Yanny}, {Lupton}, {Smith},
  {Barentine}, {Brewington}, {Brinkmann}, {Harvanek}, {Kleinman}, {Krzesinski},
  {Long}, {Nitta}, \& {Snedden}}]{zucker-etal06-cvn}
{Zucker}, D.~B. et~al. 2006{\natexlab{a}}, ApJ, 643, L103

\bibitem[{{Zucker} {et~al.}(2006{\natexlab{b}}){Zucker}, {Belokurov}, {Evans},
  {Kleyna}, {Irwin}, {Wilkinson}, {Fellhauer}, {Bramich}, {Gilmore}, {Newberg},
  {Yanny}, {Smith}, {Hewett}, {Bell}, {Rix}, {Gnedin}, {Vidrih}, {Wyse},
  {Willman}, {Grebel}, {Schneider}, {Beers}, {Kniazev}, {Barentine},
  {Brewington}, {Brinkmann}, {Harvanek}, {Kleinman}, {Krzesinski}, {Long},
  {Nitta}, \& {Snedden}}]{zucker-etal06-uma2}
{Zucker}, D.~B. et~al. 2006{\natexlab{b}}, ApJ, 650, L41

\end{thebibliography}

\end{document}